\documentclass[12pt]{article}
\usepackage{graphicx,amssymb,amsmath,empheq}
\usepackage{authblk}
\usepackage{amsthm}
\usepackage{empheq}
\usepackage{subfig} 
\usepackage{hyperref}
\hypersetup{colorlinks = true, linkcolor=black, citecolor=black, urlcolor=blue}
\usepackage{url}

\theoremstyle{plain}

\usepackage{tikz}
\usepackage{tikz-cd}
\usetikzlibrary{arrows}
\usetikzlibrary{intersections}
\usetikzlibrary{shapes.geometric}
\usetikzlibrary{decorations.pathmorphing, patterns,shapes}
\usetikzlibrary{decorations.markings}
\usepackage[margin=1.0cm]{caption}

\tikzset{
  mid arrow/.style={postaction={decorate,decoration={
        markings,
        mark=at position .575 with {\arrow[#1]{stealth}}
      }}},
  near arrow/.style={postaction={decorate,decoration={
        markings,
        mark=at position .275 with {\arrow[#1]{stealth}}
      }}},
   far arrow/.style={postaction={decorate,decoration={
        markings,
        mark=at position .800 with {\arrow[#1]{stealth}}
      }}},
}

\pgfdeclarepatternformonly{south west lines}{\pgfqpoint{-0pt}{-0pt}}{\pgfqpoint{3pt}{3pt}}{\pgfqpoint{3pt}{3pt}}{
        \pgfsetlinewidth{0.4pt}
        \pgfpathmoveto{\pgfqpoint{0pt}{0pt}}
        \pgfpathlineto{\pgfqpoint{3pt}{3pt}}
        \pgfpathmoveto{\pgfqpoint{2.8pt}{-.2pt}}
        \pgfpathlineto{\pgfqpoint{3.2pt}{.2pt}}
        \pgfpathmoveto{\pgfqpoint{-.2pt}{2.8pt}}
        \pgfpathlineto{\pgfqpoint{.2pt}{3.2pt}}
        \pgfusepath{stroke}}

\usepackage[nosort]{cite}

\newenvironment{defn}[1][Definition]{\begin{trivlist}
\item[\hskip \labelsep {\bfseries #1}]}{\end{trivlist}}

\renewcommand{\bar}{\overline}
\renewcommand{\tilde}{\widetilde}
\renewcommand{\hat}{\widehat}
\renewcommand{\leq}{\leqslant}
\renewcommand{\geq}{\geqslant}

\newcommand{\Tr}{\operatorname{Tr}}

\newcommand{\dkap}{\delta\kern-1.25pt\varkappa}

\newcommand{\calH}{\mathcal{H}}

\def\text#1{{\rm #1}}

\def\ket#1{\left|#1\right\rangle}

\def\kc#1{\left(#1\right)}

\def\ke#1{\left\{#1\right\}}

\def\trace#1{{\rm Tr}\left[#1\right]}

\def\be{\begin{equation}}       \def\ee{\end{equation}}
\def\bea{\begin{eqnarray}}      \def\eea{\end{eqnarray}}
\def\ba{\begin{array} }
\def\ea{\end{array} }

\def\=>{\Rightarrow}
\def\>{\rightarrow}

\title{Majorana fermions and the Sensitivity Conjecture
}

\author[1]{Yingfei Gu}
\affil[1]{\normalsize\it Department of Physics, Harvard University, Cambridge MA 02138, USA}
\author[2,3,4]{Xiao-Liang Qi}
\affil[2]{\normalsize\it Stanford Institute for Theoretical Physics, Stanford University, CA 94305, USA}
\affil[3]{\normalsize\it Department of Physics, Stanford University, CA 94305, USA}
\affil[4]{\normalsize\it Google, Mountain View, CA 94043, USA}

\begin{document}
  \maketitle

\abstract{Recently, Hao Huang proved the Sensitivity Conjecture, an important result about complexity measures of Boolean functions. We will discuss how this simple and elegant proof turns out to be closely related to physics concepts of the Jordan-Wigner transformation and Majorana fermions. This note is not intended to contain original results. Instead, it is a translation of the math literature in a language that is more familiar to physicists, which helps our understanding and hopefully may inspire future works along this direction.}

\section{Introduction}

We start by a brief sketch of the sensitivity conjecture and Hao Huang's proof\cite{huang2019induced}. In computer science, characterizing the complexity of Boolean functions is an important task. A Boolean function $f$ maps a bit string $s=(s_1s_2\ldots s_n)\in \{0,1\}^n$ of $n$ bits to a binary output ``$\pm 1$''.
A measure related to this purpose is called {\it sensitivity} $s(f)$, which refers to the number of bits that, if flipped, will affect the output. For example, if we define the parity function $P(s)=(-1)^{\sum_{j=1}^n s_j}$, it is $1$ if the total number of $s_j=1$ is even, and $-1$ otherwise. This function has sensitivity $n$ since flipping any bit will change the output. A more precise definition will be provided later in section~\ref{sec: sensitivity}. Here we will only provide an intuitive description. 
A generalization of sensitivity is the {\it block sensitivity} $bs(f)$, which means we group the $n$ bits into a few blocks, and allow all bits in each block to flip simultaneously. The block sensitivity refers to the maximum number of blocks that, if flipped, change the value of the function. In defining this quantity, one also maximizes over all possible partitions of blocks. Therefore one always have $bs(f)\geq s(f)$, since the latter corresponds to a particular partition with all blocks having size $1$. The question is how much larger can $bs(f)$ be compared with $s(f)$. Nisan and Szegedy \cite{nisan1994degree} proposed the sensitivity conjecture that $bs(f)\leq s(f)^C$ with $C$ some positive constant. In other words, the conjecture is that being allowed to flip blocks rather than single bits only increase the sensitivity polynomially. 

Based on preivous works by Gotsman, Linial\cite{gotsman1992equivalence} and Nisan, Szegedy\cite{nisan1994degree}, the sensitivity conjecture can be mapped to an  equivalent problem in graph theory, about subgraphs of a hypercube, which is what Huang proved in Ref. \cite{huang2019induced}. For a hypercube in $n$ dimensions, and an induced subgraph $H$ that consists of $(2^{n-1}+1)$ vertices---one more vertex than half of the hypercube, Huang proved that the maximum degree of $H$ is at least $\sqrt{n}$. Here the maximum degree of a graph is defined as the number of neighbors of a vertex $x$, maximized over all vertices.
As we will review in section~\ref{sec: sensitivity}, this result proves the sensitivity conjecture with $C=4$.

The key step in Huang's proof is to bound the degree of $H$ by maximum eigenvalue of a matrix, and the matrix is designed specially such that its eigenvalues are $\pm\sqrt{n}$. Interestingly, it turns out that the special matrix has a physical interpretation of a Majorana fermion operator. If we map each bit string to a state of $n$ qubits, they form a basis of the $2^n$ dimensional Hilbert space. For example, the $n$ qubits can be physically realized as $n$ spins in an one-dimensional quantum Ising chain. In this system, it is well-known that Majorana fermion operators can be constructed from single qubit operators by the Jordan-Wigner transformation\cite{jordan1928pauli}, which turns out to coincide exactly with Huang's construction.

In the following we will discuss the graph theory problem and Huang's proof in physicist's language in section~\ref{sec: Huang Proof} and \ref{sec: JW transformation}, and also analyze an example saturating the bound\cite{chung1988induced} in section~\ref{sec: tight}. For readers who are interested, we also reivew the relation of the graph theory problem to sensitivity conjecture in section~\ref{sec: sensitivity}. The purpose of the note is to introduce Huang's proof to physicists in their familiar language, as well as providing a physicist's view of Huang's pseudo-adjacency matrix to mathematicians. Similar observations have been recently made in terms of exterior algebra by Karasev \cite{karasev2019huang} and Clifford algebra by Tao\cite{Tao} and by Mathews\cite{mathews2019sensitivity}.

\section{Huang's theorem}
\label{sec: Huang Proof}

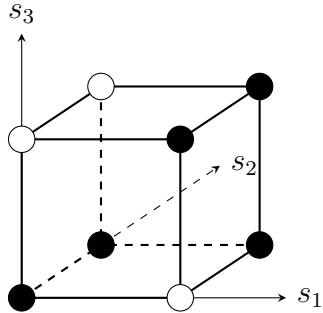
\begin{figure}[t]
    \centering
    \begin{tikzpicture}[baseline={(current bounding box.center)}]
        \draw[->,>=stealth] (0pt,0pt)--(100pt,0pt) node[right]{$s_1$};
        \draw[->,>=stealth] (0pt,0pt)--(0pt,100pt) node[above]{$s_3$};
        \draw[dashed,->,>=stealth] (0pt,0pt)--(75pt,50pt) node[right]{$s_2$};
        \draw[thick] (0pt,0pt) -- (60pt,0pt)-- (60pt,60pt)--(0pt,60pt)--(0pt,0pt);
        \draw[thick] (60pt,0pt) -- (90pt,20pt);
        \draw[thick] (60pt,60pt) -- (90pt,80pt);
        \draw[thick] (0pt,60pt) -- (30pt,80pt);
        \draw[thick,dashed] (0pt,0pt) -- (30pt,20pt);
        \draw[thick,dashed] (90pt,20pt) -- (30pt,20pt);
        \draw[thick,dashed] (30pt,80pt) -- (30pt,20pt);
        \draw[thick] (90pt,80pt) -- (90pt,20pt);
        \draw[thick] (90pt,80pt) -- (30pt,80pt);
        \filldraw[fill=black]  (0pt,0pt) circle (5pt);  
        \filldraw[fill=white]  (60pt,0pt) circle (5pt); 
        \filldraw[fill=white]  (0pt,60pt) circle (5pt); 
        \filldraw[fill=black]  (60pt,60pt) circle (5pt); 
        \filldraw[fill=black]  (30pt,20pt) circle (5pt);  
        \filldraw[fill=black]  (90pt,20pt) circle (5pt); 
        \filldraw[fill=white]  (30pt,80pt) circle (5pt); 
        \filldraw[fill=black]  (90pt,80pt) circle (5pt); 
    \end{tikzpicture}
    \caption{Black dots form a induced subgraph $H$ of $Q^3$. In $Q^3$, each vertex has degree $3$, however in the subgraph formed by the black dots, the maximum degree $\Delta(H)=2$.}
    \label{fig:3d cube}
\end{figure}

Let $Q^n$ be a $n$-dimensional hypercube graph with vertex set $\{0,1\}^n$, i.e. each vertex $s$ is represented by a bit string of length $n$, i.e. $s=(s_1s_2\ldots s_n)$ with $s_j\in \{0,1\}$. And two vertices are adjacent if and only if they differ in exactly one coordinate. An induced subgraph $H$ of $Q^n$ is another graph, formed from a subset of the vertices of $Q^n$ and all of the edges connecting pairs of vertices in that subset. 

In graph theory, the degree of a vertex of an undirected graph $G$ is the number of edges that are connected to the vertex, we denote the maximum degree over all the vertices by $\Delta(G)$. 
See figure~\ref{fig:3d cube} for an example of an induced subgraph (black dots) with $5$ vertices of the $3$-dimensional cube $Q^3$. 

Huang proved the following theorem in a recent preprint~\cite{huang2019induced} on the lower bound on maximum degree of induced subgraphs of hypercube:
\begin{defn}[Theorem 1.]
Let $H$ be an arbitrary $(2^{n-1}+1)$-vertex induced subgraph of $Q^n$, then there is a vertex in $H$ that is adjacent to at least $\sqrt{n}$ other vertices in $H$, i.e.
\begin{equation}
\Delta (H) \geq \sqrt{n} \,.\label{eq:HuangTheorem}
\end{equation}
Moreover this inequality is tight when $n$ is a perfect square.
\end{defn}

In the rest of this section, we will sketch the idea of Huang's proof of the inequality (in a slightly different way from the original proof). The tightness of the bound will be discussed later in section~\ref{sec: tight}.

\subsection{Sketch of the proof}\label{sec:sketchproof}

The key tool used in Huang's proof is a special ``pseudo-adjacency matrix". To explain this concept let us first define adjacency matrix. For an undirected graph with $D$ number of vertices, the adjacency matrix is a $D\times D$ real symmetric matrix $A^D$, with the matrix element $A^D_{st}=1$ if and only if the two vertices $s,t$ are connected by an edge, and $A^D_{st}=0$ otherwise. For example the adjacency matrix $A^{Q^n}$ of the hypercube $Q^n$ is a $2^n\times 2^n$ square matrix with the rows and columns indexed by the vertices in $Q^n$. Matrix element $A^{Q^n}_{st}=1$ if and only if the corresponding bit strings $s$ and $t$ are related by a single bit flip. Given $A^{Q^n}$, the adjacency matrix $A^H$ of the induced subgraph $H$ is a principal submatrix with rows and columns corresponding to the vertices in $H$. The maximum degree of $H$ corresponds to the maximum number of nonzero components in each row of matrix $A^H$. It is simple to show that $\Delta(H)$ for any graph is at least the largest eigenvalue of the adjacency matrix $A^H$. Therefore we can lower bound $\Delta(H)$ by the maximum eigenvalue of $A^H$.

To see how this works, let us assume $\phi$ is the eigenvector corresponding to the largest (in absolute value) eigenvalue $\lambda_1$ of $A^H$, namely $A^H \phi=\lambda_1 \phi$. Now assume $\phi_s$ has the largest absolute value among all the components in the vector $\phi$, then we have 
\begin{equation}
   | \lambda_1 \phi_s | = | (A^H\phi)_s | =\bigg| \sum_{t}  A^H_{st}\phi_t  \bigg| \leq \sum_t |A^H_{st} | |\phi_t | \leq \sum_t |A^H_{st} | |\phi_s | \leq \Delta(H) |\phi_s|\,,\label{eq:eigbound}
\end{equation}
thus we conclude $|\lambda_1|\leq \Delta(H)$. Therefore Theorem 1 will be proved if we can show $A^H$ has an eigenvalue $\sqrt{n}$. Since we do not often know the eigenvalues of the submatrix, the idea is to use the eigenvalue of the full matrix $A^{Q^n}$ to bound $A^H$. 

However, the eigenvalues of $A^{Q^n}$ include all integers $-n,-n+2,-n+4,...,+n$ (as will become clear later in the physics interpretation), so that it is difficult to prove a bound. 
A key insight is that the inequality (\ref{eq:eigbound}) also applies if we replace $A^H$ by a different matrix $\tilde{A}^H$, as long as it is entry-wise dominated by $A^H$ (i.e. $|\tilde{A}^H_{st}|\leq A^H_{st},~\forall s,t$). Among such matrices, Huang found a ``magic" Hermitian matrix $\tilde{A}^{Q^n}$ which satisfies the following properties:
\begin{equation}
    \left|\tilde{A}^{Q^n}_{st}\right|=A^{Q^n}_{st} \,, \quad \Tr \tilde{A}^{Q^n}=0 \,, \quad \tilde{A}^{Q^n\dagger}\tilde{A}^{Q^n}=n\mathbb{I}\,.
    \label{eqn: A tilde}
\end{equation}
The second and third equation above imply that $\tilde{A}^{Q^n}$ have only two eigenvalues $\pm \sqrt{n}$, and each eigenvalue has degeneracy $2^{n-1}$. Then we can define the subspace spanned by its positive eigenvalues as $E_+$. In other words, $E_+$ is the $2^{n-1}$ dimensional space of all vectors satisfying $\tilde{A}^{Q^n}\phi=\sqrt{n}\phi$. On the other hand, the subgraph $H$ also corresponds to a subspace $V_H$, spanned by the vertices in $H$, which has dimension $(2^{n-1}+1)$. By dimension counting, $V_H$ and $E_+$ must have a nonzero intersection. That is to say, there must be a vector $\phi_H$ in $V_H$ which is also in $E_+$, such that $\tilde{A}^{H}\phi_H=\tilde{A}^{Q^n}\phi_H=\sqrt{n}\phi_H$. Therefore we proved that the induced submatrix $\tilde{A}^H$ has an eigenvalue $\sqrt{n}$, such that $\Delta(H)\geq \sqrt{n}$.

\begin{defn}[Huang's construction of the matrix $\tilde{A}^{Q^n}$.] Here we summarize Huang's construction of $\tilde{A}^{Q^n}$, before explaining its physical interpretation in next subsection. The matrix is defined iteratively as follows:
\begin{equation}
A_1 = \begin{pmatrix}
0 & 1\\
1 & 0
\end{pmatrix}\,, \quad A_m= 
 \begin{pmatrix}
A_{m-1} & \mathbb{I}_{2^{m-1}}\\
\mathbb{I}_{2^{m-1}} & -A_{m-1}
\end{pmatrix} \,.
\label{An matrix}
\end{equation}
and $\tilde{A}^{Q^n}=A_n$. One can check that
$A_n^2=n\mathbb{I}$ and $\Tr(A_n)=0$.
\end{defn}

\section{Pseudo-adjacency matrix from Majorana fermions}
\label{sec: JW transformation}

From the first glance, it is unclear why the pseudo-adjacency matrix $A_n$ constructed by Huang is special, compared to other choices of sign (or phase). Interestingly, this matrix has a very simple interpretation if we relate the hypercube to a quantum physics problem of spin chains. In this section, we will provide an overview of the spin chain problem and the Jordan-Wigner transformation which maps spins to Majorana fermions, and show how Huang's pseudo-adjacency matrix is simply the momentum zero component of Majorana fermion operator.

\subsection{Spin chain}
\label{subsec: pauli}

\begin{figure}[t]
    \centering 
    \begin{tikzpicture}[scale=1.2,baseline={(current bounding box.center)}]
    \draw (0pt,0pt) -- (140pt,0pt);
    \filldraw[fill=black] (0pt,0pt) circle (3pt);
    \draw[->,>=stealth,thick] (0pt,8pt)--(0pt,-8pt) node[below]{ $1$};
    \filldraw[fill=black] (20pt,0pt) circle (3pt);
    \draw[<-,>=stealth,thick] (20pt,8pt)--(20pt,-8pt) node[below]{ $0$};
    \filldraw[fill=black] (40pt,0pt) circle (3pt);
    \draw[->,>=stealth,thick] (40pt,8pt)--(40pt,-8pt) node[below]{ $1$};
    \filldraw[fill=black] (60pt,0pt) circle (3pt);
    \draw[<-,>=stealth,thick] (60pt,8pt)--(60pt,-8pt) node[below]{ $0$};
    \filldraw[fill=black] (80pt,0pt) circle (3pt);
    \draw[<-,>=stealth,thick] (80pt,8pt)--(80pt,-8pt) node[below]{ $0$};
    \filldraw[fill=black] (100pt,0pt) circle (3pt);
    \draw[->,>=stealth,thick] (100pt,8pt)--(100pt,-8pt) node[below]{ $1$};
    \filldraw[fill=black] (120pt,0pt) circle (3pt);
    \draw[->,>=stealth,thick] (120pt,8pt)--(120pt,-8pt) node[below]{ $1$};
    \filldraw[fill=black] (140pt,0pt) circle (3pt);
    \draw[<-,>=stealth,thick] (140pt,8pt)--(140pt,-8pt) node[below]{ $0$};
    \end{tikzpicture}
    \caption{A spin configuration $|\! \downarrow \uparrow \downarrow\uparrow\uparrow\downarrow\downarrow \uparrow \rangle$ of $8$ spins. It can also be denoted by a bit string $|10100110\rangle$.}
    \label{fig:spin chain}
\end{figure}
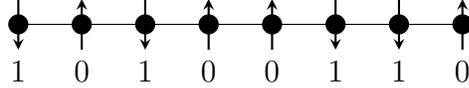

The bit string of length $n$ can be naturally mapped to a basis of quantum states in a spin chain of the same length. Physically, a quantum spin chain is realized by a system with $n$ atoms, each has an electron with spin $1/2$ (plus possibly other electrons which form spin singlet and thus can be neglected in the discussion here). If the electrons cannot move in space, for example due to Coulomb repulsion, the spin degree of freedom is the only one we need to consider (see Fig.~\ref{fig:spin chain}.) In that case, the quantum state of the system lives in a $2^n$ dimensional Hilbert space $\calH=\bigotimes_{j=1}^n\calH_j$. A basis of the Hilbert space can be defined by the eigenstates of a spin component, such as the $z$-component. If we denote the two spin states $S_z=+1/2$, $S_z=-1/2$ on each site by binary values $\ket{0}$, $\ket{1}$ respectively, each length-$n$ bit string $s=\ke{s_1s_2...s_n}$ corresponds to a spin eigenstate $\ket{s}=\ket{s_1}\otimes\ket{s_2}\otimes...\otimes\ket{s_n}$. 

It is convenient to use the Pauli matrices $X,Y,Z$ which are a set of three $2\times 2$ Hermitian and unitary matrices
\begin{equation}
    X=\begin{pmatrix}
    0 & 1\\
    1 & 0
    \end{pmatrix}\,, \quad 
    Y=\begin{pmatrix}
    0 & -i\\
    i & 0
    \end{pmatrix}\,, \quad
    Z=\begin{pmatrix}
    1 & 0\\
    0 & -1
    \end{pmatrix} \,.
\end{equation}
They act on a $2$ dimensional Hilbert space $\calH$ of a spin $1/2$ particle and have the physical meaning of twice the spin components along $x$, $y$ and $z$ direction respectively. We will also denote $X_j,Y_j,Z_j$ as the Pauli operators acting on $j$-th spin, such that in the entire Hilbert space $X_j=\mathbb{I}_{2^{j-1}}\otimes X\otimes\mathbb{I}_{2^{n-j}}$ and similar for $Y_j,Z_j$.  

The bit string notation suggests a correspondence between the quantum system of $n$ spins and the hypercube $Q^n$. Each vertex of $Q^n$ corresponds to a basis vector of the Hilbert space $\calH$. Furthermore, two adjacent vertices in $Q^n$ correspond to two spin configurations that differ by exactly one spin flip. Note that the Pauli matrix $X_j$ acts on spin configurations by flipping their $j$-th spin, one can see that the adjacency matrix of the hypercube can be written as
\begin{equation}
    A^{Q^n}=\sum_{j=1}^nX_j\,,
    \label{eq:adjX}
\end{equation}
since the right-hand side operator has a nonzero matrix element between two basis states $\ket{s},\ket{t}$ if and only if they are related by one of the $X_j$, and are therefore related by a single bit flip on site $j$. In other words, the adjacency matrix of the hypercube is nothing but the net spin along the $x$-direction. Since $X_j$ all commute with each other, we know that the eigenvalues of $A^{Q^n}$ are integers $-n,-n+2,-n+4,...,+n$.

\subsection{Jordan-Wigner transformation}

Compared with the adjacency matrix in Eq.~\eqref{eq:adjX}, the key difference of the pseudo-adjacency matrix used by Huang is that it is a sum of {\it anti-commuting} terms rather than commuting ones. This turns out to be related to a known transformation in physics, named as the Jordan-Wigner transformation, which maps the spin chain to a fermion chain. Some spin models such as the quantum Ising model can be solved by the Jordan-Wigner transformation since the corresponding fermion problem is simple.

The Jordan-Wigner transformation maps spin operators onto Majorana operators as follows (See. Fig.~\ref{fig:JW transformation}), 
\begin{equation}
\psi_j = X_j \prod_{k=j+1}^n Z_k\,, \quad
\eta_j = Y_j \prod_{k=j+1}^n Z_k\,.
\end{equation}
The Majorana operators $\psi_j$ and $\eta_j$ are Hermitian and obey the following anti-commutation relation
\begin{equation}
    \{ \psi_j,\psi_k\} = \{ \eta_j,\eta_k\}  = 2 \delta_{jk} \,, \quad  \{ \psi_j,\eta_k\} =0 \,.
\end{equation}
The anti-commutation relation originates from that of Pauli operators. Mathematically, $\psi_j, \eta_j$ are generators of a Clifford algebra. 

We now make a few comments on the Jordan-Wigner transformation
\begin{figure}
    \centering
   \begin{tikzpicture}[scale=1.2]
        \node at (-15pt,5pt) {$\psi_2=$};
        \draw (5pt,5pt)--(140pt,5pt);
        \filldraw[fill=white] (0pt,0pt) rectangle ++(10pt,10pt);
        \node at (5pt,5pt) {\scriptsize $I$};
        \filldraw[fill=white] (20pt,0pt) rectangle ++(10pt,10pt);
        \node at (25pt,5pt) {\scriptsize $X$};
        \filldraw[fill=white] (40pt,0pt) rectangle ++(10pt,10pt);
        \node at (45pt,5pt) {\scriptsize $Z$};
        \filldraw[fill=white] (60pt,0pt) rectangle ++(10pt,10pt);
        \node at (65pt,5pt) {\scriptsize $Z$};
        \filldraw[fill=white] (80pt,0pt) rectangle ++(10pt,10pt);
        \node at (85pt,5pt) {\scriptsize $Z$};
        \filldraw[fill=white] (100pt,0pt) rectangle ++(10pt,10pt);
        \node at (105pt,5pt) {\scriptsize $Z$};
        \filldraw[fill=white] (120pt,0pt) rectangle ++(10pt,10pt);
        \node at (125pt,5pt) {\scriptsize $Z$};
        \filldraw[fill=white] (140pt,0pt) rectangle ++(10pt,10pt);
        \node at (145pt,5pt) {\scriptsize $Z$};
        
        \node at (-15pt,-15pt) {$\psi_6=$};
        \draw (5pt,-15pt)--(140pt,-15pt);
        \filldraw[fill=white] (0pt,-20pt) rectangle ++(10pt,10pt);
        \node at (5pt,-15pt) {\scriptsize $I$};
        \filldraw[fill=white] (20pt,-20pt) rectangle ++(10pt,10pt);
        \node at (25pt,-15pt) {\scriptsize $I$};
        \filldraw[fill=white] (40pt,-20pt) rectangle ++(10pt,10pt);
        \node at (45pt,-15pt) {\scriptsize $I$};
        \filldraw[fill=white] (60pt,-20pt) rectangle ++(10pt,10pt);
        \node at (65pt,-15pt) {\scriptsize $I$};
        \filldraw[fill=white] (80pt,-20pt) rectangle ++(10pt,10pt);
        \node at (85pt,-15pt) {\scriptsize $I$};
        \filldraw[fill=white] (100pt,-20pt) rectangle ++(10pt,10pt);
        \node at (105pt,-15pt) {\scriptsize $X$};
        \filldraw[fill=white] (120pt,-20pt) rectangle ++(10pt,10pt);
        \node at (125pt,-15pt) {\scriptsize $Z$};
        \filldraw[fill=white] (140pt,-20pt) rectangle ++(10pt,10pt);
        \node at (145pt,-15pt) {\scriptsize $Z$};
        
        \node at (-15pt,-35pt) {$\eta_5=$};
        \draw (5pt,-35pt)--(140pt,-35pt);
        \filldraw[fill=white] (0pt,-40pt) rectangle ++(10pt,10pt);
        \node at (5pt,-35pt) {\scriptsize $I$};
        \filldraw[fill=white] (20pt,-40pt) rectangle ++(10pt,10pt);
        \node at (25pt,-35pt) {\scriptsize $I$};
        \filldraw[fill=white] (40pt,-40pt) rectangle ++(10pt,10pt);
        \node at (45pt,-35pt) {\scriptsize $I$};
        \filldraw[fill=white] (60pt,-40pt) rectangle ++(10pt,10pt);
        \node at (65pt,-35pt) {\scriptsize $I$};
        \filldraw[fill=white] (80pt,-40pt) rectangle ++(10pt,10pt);
        \node at (85pt,-35pt) {\scriptsize $Y$};
        \filldraw[fill=white] (100pt,-40pt) rectangle ++(10pt,10pt);
        \node at (105pt,-35pt) {\scriptsize $Z$};
        \filldraw[fill=white] (120pt,-40pt) rectangle ++(10pt,10pt);
        \node at (125pt,-35pt) {\scriptsize $Z$};
        \filldraw[fill=white] (140pt,-40pt) rectangle ++(10pt,10pt);
        \node at (145pt,-35pt) {\scriptsize $Z$};
   \end{tikzpicture}
    \caption{Three examples of the Jordan-Wigner transformation: $\psi_2$, $\psi_6$ and $\eta_5$ for a system with $8$ spins. Comparing to their bosonic counterpart $X_2$, $X_6$ and $Y_5$, the fermionic operators have strings of $Z$ operator attached to them from the right end. We refer such strings as the Jordan-Wigner strings.}
    \label{fig:JW transformation}
\end{figure}
\begin{enumerate}
    \item The Majorana operators $\psi_j$ and $\eta_j$ may be viewed as the spin operator $X_j$ and $Y_j$ attached to a Jordan-Wigner string $\Pi_{k=j+1}^n Z_k$ as shown in figure~\ref{fig:JW transformation}. The definition of the transformation is not unique, e.g. one may alternatively choose to attach a string starting from the left, i.e. $\Pi_{k=1}^{j-1}Z_k$;
    \item 
    Using fermionic operators $\psi_j$ and $\eta_j$, one can define the complex fermion annihilation and creation operators $c,c^\dagger$ as follows, 
    \begin{equation}
    \begin{aligned}
    &c_j=\frac{1}{2} \left( \psi_j + i \eta_j \right)\,, \quad c^\dagger_j=\frac{1}{2} \left( \psi_j - i \eta_j \right)\,, \\
    &~\{c_j,c_k \}=\{c_j^\dagger,c_k^\dagger\} = 0 \,, \quad \{c_j,c^\dagger_k\}=\delta_{jk}\,.
    \end{aligned}
    \end{equation}
    The fermion number on site $j$ is $n_j=c_j^\dagger c_j$, which can be written in spin operators as $n_j=\kc{1-Z_j}/2$. Therefore the two states with fermion number $0,1$ corresponds to spin $Z=+1,-1$ respectively. 
    Up to some sign convention, the fermion number eigenstate basis is identical to the spin $Z$ basis states, which corresponds to the vertices of the hypercube. 
\end{enumerate}

\subsection{Pseudo-adjacency matrix}

The Majorana fermion operator $\psi_j$ flips the value of the $j$-th bit in the same way as the Pauli operator $X_j$, except that there is a fermion sign which depends on the bit string in a non-local way. Therefore if we define
\begin{equation}
    \tilde{A}=\sum_{j=1}^n\psi_j\,,
\end{equation}
it is easy to prove that $\tilde{A}$ is a pseudo-adjacency matrix. Actually for each matrix element, $\big|\tilde{A}_{st}\big|=\big|A^{Q^n}_{st}\big|$. Physically, $\tilde{A}$ is related to the Fourier transform of the Majorana fermion at momentum $p=0$:
\begin{equation}
    \gamma_p=\frac1{\sqrt{n}}\sum_{j=1}^n\psi_je^{-ipj}\label{eq:ftMajorana}\,,
\end{equation}
which satisfies $\{\gamma_p,\gamma_q^\dagger\}=2\delta_{pq}$ for momentum $p=\frac{2\pi}{n}k,k=0,1,...,n-1$. In particular, $\gamma_0=\frac1{\sqrt{n}}\tilde{A}$ is a Majorana operator satisfying $\gamma_0^2=\mathbb{I},~\trace{\gamma_0}=0$. Therefore we see that $\tilde{A}$ satisfies the conditions in \eqref{eqn: A tilde}. 
Actually, $\tilde{A}$ is exactly Huang's pseudo-adjacency matrix defined in Eq.~(\ref{An matrix}). The important value $\sqrt{n}$ which appears in the eigenvalue of $\tilde{A}$ is actually the normalization factor in the Fourier transform (\ref{eq:ftMajorana}). 

In summary, we have shown that the pseudo-adjacency matrix is a uniform superposition of Majorana operators, and the important feature of large spectrum gap is due to anti-commutation relation of fermions. As a consequence of this Majorana fermion interpretation, we see that there is a continuous family of other pseudo-adjacency matrices that work equally well. On each site we can define a superposition of the two Majorana fermions:
\begin{equation}
    \chi_j=\cos\theta_j\psi_j+\sin\theta_j \eta_j=\kc{\cos\theta_j X_j+\sin\theta_j Y_j}\prod_{k=j+1}^nZ_k 
\end{equation}
with arbitrary site-dependent angle variables $\theta_j$. $\chi_j$ satisfies the anti-commutation relation $\ke{\chi_j,\chi_k}=2\delta_{jk}$. It also does the same job of flipping $j$-th spin, but with a phase:
\begin{equation}
    \chi_j\ket{0}_j=\pm e^{i\theta_j}\ket{1}_j\,, \quad \chi_j\ket{1}_j=\pm e^{-i\theta_j}\ket{0}_j \,.
\end{equation}
Here the $\pm$ sign depends on the value of other bits. Therefore if we define
\begin{equation}
    A^\theta=\sum_{j=1}^n\chi_j\,,
\end{equation}
it is easy to verify that $A^\theta$ satisfies the conditions in Eq.~\eqref{eqn: A tilde}.

\section{Saturating the lower bound}
\label{sec: tight}

The lower bound $\sqrt{n}$ that Huang proved can not be improved (except the trivial improvement to the ceiling $\lceil \sqrt{n} \rceil$ for an integer $n$ that is not a perfect square). An example of $\Delta(H)=\sqrt{n}$ was shown by Chung, Furedi, Graham and Seymour\cite{chung1988induced}, although it wasn't known to be the lower bound. (See also  \cite{Tao} for further discussion on this example.) The spin model picture is helpful in understanding this example, so we summarize it in this section. In addition to defining the subgraph $H$, we also explicitly construct the eigenvector of $\tilde{A}^H$ with eigenvalue $\sqrt{n}$ in the form of a quantum state.

For convenience, we will assume $n=l^2$ is a perfect square in this section. 

\subsection{The construction of Chung, Furedi, Graham and Seymour}
\label{subsec: CFGS example}

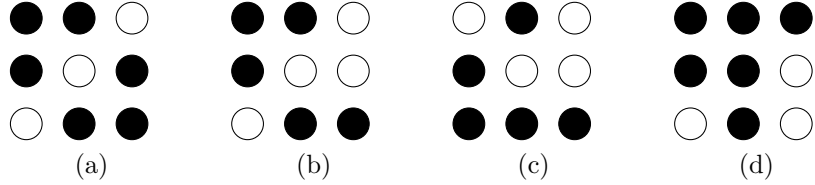
\begin{figure}
    \centering
    \subfloat[]{
    \begin{tikzpicture}[scale=1,baseline={(current bounding box.center)}]
    \filldraw[fill=white] (0pt,0pt) circle (6pt) ; 
    \filldraw[fill=black] (20pt,0pt) circle (6pt);
    \filldraw[fill=white] (20pt,20pt) circle (6pt);
    \filldraw[fill=black] (0pt,20pt) circle (6pt);
     \filldraw[fill=black] (40pt,0pt) circle (6pt); 
     \filldraw[fill=black] (0pt,40pt) circle (6pt);
    \filldraw[fill=black] (20pt,40pt) circle (6pt);
    \filldraw[fill=black] (40pt,20pt) circle (6pt);
    \filldraw[fill=white] (40pt,40pt) circle (6pt);
    \end{tikzpicture}
    }
    \qquad
    \subfloat[]{
    \begin{tikzpicture}[scale=1,baseline={(current bounding box.center)}]
    \filldraw[fill=white] (0pt,0pt) circle (6pt); 
    \filldraw[fill=black] (20pt,0pt) circle (6pt);
    \filldraw[fill=white] (20pt,20pt) circle (6pt);
    \filldraw[fill=black] (0pt,20pt) circle (6pt);
     \filldraw[fill=black] (40pt,0pt) circle (6pt);
     \filldraw[fill=black] (0pt,40pt) circle (6pt);
    \filldraw[fill=black] (20pt,40pt) circle (6pt);
    \filldraw[fill=white] (40pt,20pt) circle (6pt);
    \filldraw[fill=white] (40pt,40pt) circle (6pt);
    \end{tikzpicture}
    }
    \qquad
    \subfloat[]{
    \begin{tikzpicture}[scale=1,baseline={(current bounding box.center)}]
    \filldraw[fill=black] (0pt,0pt) circle (6pt);
    \filldraw[fill=black] (20pt,0pt) circle (6pt);
    \filldraw[fill=white] (20pt,20pt) circle (6pt);
    \filldraw[fill=black] (0pt,20pt) circle (6pt);
     \filldraw[fill=black] (40pt,0pt) circle (6pt);
     \filldraw[fill=white] (0pt,40pt) circle (6pt);
    \filldraw[fill=black] (20pt,40pt) circle (6pt);
    \filldraw[fill=white] (40pt,20pt) circle (6pt);
    \filldraw[fill=white] (40pt,40pt) circle (6pt);
    \end{tikzpicture}
    }
    \qquad
    \subfloat[]{
    \begin{tikzpicture}[scale=1,baseline={(current bounding box.center)}]
    \filldraw[fill=white] (0pt,0pt) circle (6pt);
    \filldraw[fill=black] (20pt,0pt) circle (6pt);
    \filldraw[fill=black] (20pt,20pt) circle (6pt);
    \filldraw[fill=black] (0pt,20pt) circle (6pt);
     \filldraw[fill=white] (40pt,0pt) circle (6pt);
     \filldraw[fill=black] (0pt,40pt) circle (6pt);
    \filldraw[fill=black] (20pt,40pt) circle (6pt);
    \filldraw[fill=white] (40pt,20pt) circle (6pt);
    \filldraw[fill=black] (40pt,40pt) circle (6pt);
    \end{tikzpicture}
    }
    \caption{Examples of bit string configurations drawn on the chessboard. According to the definition, (a) and (b) belongs to $U$, while (c) and (d) belongs to its complement $\bar{U}$.}
    \label{fig:l by l square}
\end{figure}

Defining an induced subgraph $H$ of $Q^n$ is equivalent to defining a Boolean function $g: \{0,1\}^n \rightarrow \{-1,1\}$, such that $g(s)=1$ if $s\in H$, and $g(s)=-1$ otherwise.\footnote{It should be noted that the maximum degree of $H$ does not tell us about the sensitivity of $g(s)$. The relation between the graph property (\ref{eq:HuangTheorem}) of a subgraph $H$ and sensitivity of Boolean functions is more complicated, which will be reviewed in next section.} 
When $n=l^2$ is a perfect square, it is convenient to order the sites into a two-dimensional lattice, as is shown in Fig.~\ref{fig:l by l square}. Correspondingly, we label the sites by two coordinates $\alpha$ (row) and $\beta$ (column) with $\alpha,\beta=1,2,...,l$. On each site $(\alpha\beta)$, the two values $s_{\alpha\beta} =0,1$ corresponds to two eigenstates of 
\begin{equation}
    n_{\alpha\beta}=c_{\alpha\beta}^\dagger c_{\alpha\beta}=\frac12\kc{1-Z_{\alpha\beta}}\,.
\end{equation}
The function $g(s)$ can always be expressed in term of $n_{\alpha\beta}$. We define the following $g(s)$ together with an auxiliary function $h(s)$, before providing its interpretation:
\begin{equation}
    \quad g(s) = P(s)h(s)\,, \qquad h(s)= 2 \prod_{\alpha=1}^l \Big( 1-\prod_{\beta=1}^ln_{\alpha\beta} \Big) -1   \,,
    \label{eqn: g and h}
\end{equation}
with $P(s)=\prod_{\alpha,\beta=1}^{l}Z_{\alpha\beta}$ the parity function. 

From the definition we see that $h(s)=1$ if and only if all rows $\alpha=1,2,...l$ satisfy $\prod_{\beta=1}^ln_{\alpha\beta}=0$, which means that each row has at least one bit with value $0$. We denote the subset with $h(s)=1$ as $U$, which is illustrated in Fig.~\ref{fig:l by l square}. $g(s)$ is defined to be equal to $h(s)$ for parity even states with $P(s)=1$, and different from $h(s)$ for parity odd states. Physically, the subgraph $H$ defined by $g(s)=1$ contains the following fermion configurations:
\begin{itemize}
    \item Each row has at least one $0$, while the total fermion number is even;
    \item At least one of the rows have all $1$'s, while the total fermion number is odd.
\end{itemize}

Now we consider the action of pseudo-adjacency matrix $\tilde{A}=\sum_{\alpha\beta}\psi_{\alpha\beta}$, which creates or annihilates one of the fermions. Since $\tilde{A}$ changes fermion number parity, it can only couple between the even and odd parity states in $H$. Let us look at an odd parity state $\ket{s}$ with $s\in H$, with at least one row of all $1$'s. In order for $\tilde{A}$ to have matrix element between $\ket{s}$ and another states $\ket{t},t\in H$, $s$ has to only contain a single row with all ones, and one of those bits must be flipped by $\tilde{A}$. Therefore there are only $l=\sqrt{n}$ ways to do that. Similarly, for an even parity state $\ket{s'}\in H$, with all rows containing at least one $0$, in order for $\tilde{A}$ to have matrix element between $\ket{s'}$ and another states $\ket{t'},t'\in H$,
 there has to be at least one row with exactly one $0$, such that $\tilde{A}$ can fill this hole. The vertex $t'$ that has most neighbors are those with exactly one $0$ in each row, for which $\tilde{A}$ can bring it to a superposition of $l=\sqrt{n}$ states in $H$. Therefore the maximum degree of $H$ is shown to be $l=\sqrt{n}$. 

For a general two-dimensional layout of $n$ (not necessarily a perfect square) sites, we denote the length of each row by $l_\alpha$, with $\alpha=1,2,\ldots,r$ where $r$ is the total number of rows. Note $n=\sum_{\alpha=1}^r l_\alpha \leq r \max_{\alpha} l_\alpha $. 
Following the above discussion, we conclude that 
the maximum degree is determined by the larger value in the maximum length of all rows and the number of rows, namely
\begin{equation}
    \Delta(H) = \max \left( \max_{\alpha} l_\alpha ,r \right) \geq \lceil \sqrt{n} \rceil \,.
\end{equation}
We remind that this inequality is stated in the specific construction of $H$ discussed above and should not be confused with Theorem 1 which applies to all possible protocols. The point here is that the optimal value $\lceil \sqrt{n} \rceil$ is achievable in this specific construction by constraining the layout into a $\lceil \sqrt{n} \rceil$ by $\lceil \sqrt{n} \rceil$ frame.

The remaining task is to compute the size of $H$, i.e. the number of vertices in $H$. It is convenient to translate this task to taking a trace of an operator. If we view fermion number $n_{\alpha\beta}$ as operators, $g(s)$ is a diagonal matrix, namely it corresponds to a quantum operator $\hat{g}=\sum_s g(s)|s\rangle \langle s |$. By abuse of notation, we denote the number of vertices in $H$ (i.e. those vertices $s$ with $g(s)=1$) by $|H|$, we have
\begin{equation}
    |H| = \sum_{s\in \{0,1\}^n} \frac{g(s)+1}{2} = \frac{1}{2} \Tr \left( 1 + g(s)
    \right) \,.
\end{equation}
To compute this quantity we expand $h(s)$ in \eqref{eqn: g and h} as monomials of $Z_{\alpha\beta}$:
\begin{equation}
    h(s)= 2 \prod_{\alpha=1}^l \left[ 1-2^{-l} \prod_{\beta=1}^l(1-Z_{\alpha\beta}) \right] -1 = a_0+\sum_{\alpha\beta}a_1^{\alpha\beta}Z_{\alpha\beta}+...+a_n\prod_{\alpha,\beta=1}^l Z_{\alpha\beta}\,.
    \label{eqn: monomial expansion}
\end{equation}
In the trace $\Tr(g(s))=\Tr(h(s)P(s))$, only the last term in the expansion \eqref{eqn: monomial expansion} contributes. Expanding the equation above we obtain
\begin{equation}
    \trace{h(s)P(s)}=2^{n}a_n=2(-1)^{l(l+1)}=2 \,.
\end{equation}
Thus, 
\begin{equation}
    |H| = \frac{1}{2} \Tr \left( 1 + g(s) \right) = 2^{n-1} + 1 \,.
\end{equation}
Therefore $H$ is an example of a subgraph with $(2^{n-1}+1)$ vertices, and maximum degree $\sqrt{n}$.

\subsection{Construction of an eigenvector}

We have shown that the equal sign in Theorem 1 can be achieved by the $H$ constructed in the above subsection. Moreover, there should be an eigenvector of the pseudo-adjacency $\tilde{A}^{Q^n}$ with eigenvalue $l=\sqrt{n}$ and supports on the subspace $V_H$. We would like to construct such eigenvector using the fermion creation and annihilation operators we defined before. The construction is partially motivated by a remark in Tao's blog\cite{Tao}.

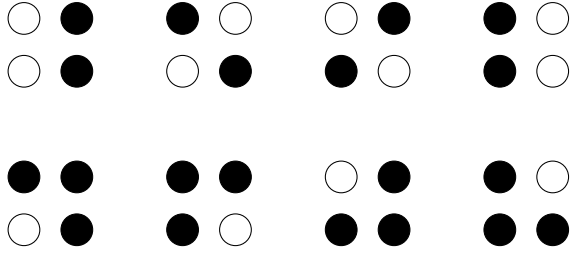
\begin{figure}[t]
    \centering
    \begin{tikzpicture}
    \filldraw[fill=white] (0pt,0pt) circle (6pt);
    \filldraw[] (20pt,0pt) circle (6pt);
    \filldraw[] (0pt,20pt) circle (6pt);
    \filldraw[] (20pt,20pt) circle (6pt);
    
    \filldraw[] (60pt,0pt) circle (6pt);
    \filldraw[fill=white] (80pt,0pt) circle (6pt);
    \filldraw[] (60pt,20pt) circle (6pt);
    \filldraw[] (80pt,20pt) circle (6pt);
    
    \filldraw[] (120pt,0pt) circle (6pt);
    \filldraw[] (140pt,0pt) circle (6pt);
    \filldraw[fill=white] (120pt,20pt) circle (6pt);
    \filldraw[] (140pt,20pt) circle (6pt);
    
    \filldraw[] (180pt,0pt) circle (6pt);
    \filldraw[] (200pt,0pt) circle (6pt);
    \filldraw[] (180pt,20pt) circle (6pt);
    \filldraw[fill=white] (200pt,20pt) circle (6pt);
    
    \filldraw[fill=white] (0pt,60pt) circle (6pt);
    \filldraw[] (20pt,60pt) circle (6pt);
    \filldraw[fill=white] (0pt,80pt) circle (6pt);
    \filldraw[] (20pt,80pt) circle (6pt);
    
    \filldraw[fill=white] (60pt,60pt) circle (6pt);
    \filldraw[] (80pt,60pt) circle (6pt);
    \filldraw[] (60pt,80pt) circle (6pt);
    \filldraw[fill=white] (80pt,80pt) circle (6pt);
    
    \filldraw[] (120pt,60pt) circle (6pt);
    \filldraw[fill=white] (140pt,60pt) circle (6pt);
    \filldraw[fill=white] (120pt,80pt) circle (6pt);
    \filldraw[] (140pt,80pt) circle (6pt);
    
    \filldraw[] (180pt,60pt) circle (6pt);
    \filldraw[fill=white] (200pt,60pt) circle (6pt);
    \filldraw[] (180pt,80pt) circle (6pt);
    \filldraw[fill=white] (200pt,80pt) circle (6pt);
    \end{tikzpicture}
    \caption{Non-zero components in $|\psi\rangle$ for $l=2$.}
    \label{fig:eigenvector}
\end{figure}

Let us still use the $l$ by $l$ lattice introduced in the last subsection and assign fermionic occupation basis to each site. For convenience, let us use the following notation
\begin{equation}
    B_\alpha = \sum_{\beta=1}^l c_{\alpha\beta}\,, \quad \{ B_\alpha,B_{\alpha'}^\dagger \} = l  \delta_{\alpha \alpha'} \,.
\end{equation}
Physically, $B_\alpha$ is proportional to the zero-momentum fermion annihilation operator on $\alpha$-th row. Now we construct a state as follows
\begin{equation}
|\psi \rangle = \left( 1 + \frac1l{\sum_{\alpha=1}^l B_\alpha^\dagger } \right) \prod_{\alpha=1}^l B_\alpha|1\rangle\,,
\end{equation}
where $|1\rangle$ is the state of ``all $1$" string, which corresponds to all fermions occupied. Physically, $\prod_{\alpha=1}^l B_\alpha |1\rangle$ is a state with $l^2-l$ fermions. If we use momentum basis in horizontal direction and coordinate basis in vertical direction, all fermion states are occupied except those with horizon momentum $0$. Then the operator ${\sum_{\alpha=1}^l B_\alpha^\dagger}$ creates one of the $l$ missing fermions. Notice that $l^2-l$ is always even, one can check that $\ket{\psi}$ only consists of terms satisfying the condition of subset $H$ defined in the previous subsection. In other words, $\ket{\psi}\in V_H$. 

Note that the pseudo-adjacency matrix $\tilde{A}$ can be expressed in terms of $B_J$ as follows,
\begin{equation}
    \tilde{A} = \sum_{\alpha=1}^{l} \left( B_\alpha + B^\dagger_\alpha \right) \,.
\end{equation}
We claim $|\psi \rangle$ is an eigenvector of $\tilde{A}$ with eigenvalue $l$. Indeed,
\begin{equation}
\begin{aligned}
    \tilde{A} | \psi \rangle &= \sum_{\alpha=1}^{l} \left( B_\alpha + B^\dagger_\alpha  \right)\left( 1 + \frac{\sum_{\alpha=1}^l B_\alpha^\dagger}{l} \right) \prod_{\alpha=1}^l B_\alpha |1\rangle \\
    &= \left( \sum_{\alpha=1}^{l} B^\dagger_\alpha + \frac{1}{l} \left\{ \sum_{\alpha=1}^l B_\alpha, \sum_{\alpha=1}^l B_{\alpha}^\dagger \right\} \right)\prod_{\alpha=1}^l B_\alpha | 1 \rangle = l | \psi \rangle \,.
\end{aligned}
\end{equation}
In writing this equation we have used $B_\alpha^2={B_\alpha^\dagger}^2=0$. 

One can also verify that when we expand $|\psi \rangle$ in the bit string basis $\ket{s}, s\in H$, its wavefunction on each basis vector $\ket{s}$ has equal absolute value, which is a requirement for the bound to be saturated. Figure~\ref{fig:eigenvector} illustrated the configurations in $\ket{\psi}$ for $l=2$.

\section{Relation to sensitivity}
\label{sec: sensitivity}

In this section, we provide some more background introduction of the relation between the graph theory problem (Theorem $1$) and the sensitivity of Boolean functions. More specifically, we would like to explain an argument by Gotsman and Linial\cite{gotsman1992equivalence} that relates Theorem 1 to a lower bound of the sensitivity and make a few comments related to physics.

We have discussed the concept of sensitivity in the introduction, but we would like to give a more precise definition here.

\begin{defn}[Sensitivity.]
For a Boolean function $f$: $\{0,1\}^n \rightarrow \{ -1,1\}$, the local sensitivity $s(f,x)$ on the input $x$ is defined as the number of coordinate $x_j$, such that $f(x)= -f(x')$ where $x'$ and $x$ differ in exactly the coordinate $x_j$. And the (global) sensitivity $s(f)=\max_x s(f,x)$ is the maximum over all inputs. 
\end{defn}

Each Boolean function $f(s)$ corresponds to a induced subgraph $G$ of the hypercube defined by $f(s)=1$. In the hypercube picture, The local sensitivity $s(f,x)$ is the number of neighbors of $x$ that is not in $G$. 

\begin{figure}[t]
    \centering
    \subfloat[Degree]{
    \begin{tikzpicture}[scale=1]
    \draw (0pt,0pt) -- (140pt,0pt);
    \filldraw[fill=black] (0pt,0pt) circle (3pt);
    \filldraw[fill=black] (20pt,0pt) circle (3pt);
    \filldraw[fill=black] (40pt,0pt) circle (3pt);
    \filldraw[fill=black] (60pt,0pt) circle (3pt);
    \filldraw[fill=black] (80pt,0pt) circle (3pt);
    \filldraw[fill=black] (100pt,0pt) circle (3pt);
    \filldraw[fill=black] (120pt,0pt) circle (3pt);
    \filldraw[fill=black] (140pt,0pt) circle (3pt);
    \draw[dashed] (02pt,8pt) -- (25pt,30pt);
    \draw[dashed] (20pt,8pt) -- (25pt,30pt);
    \draw[dashed] (56pt,8pt) -- (25pt,30pt);
    \filldraw[fill=black] (25pt,30pt) circle (1pt) node[above]{\scriptsize $Z_1Z_2Z_4$};
    \draw[dashed] (46pt,8pt) -- (90pt,50pt);
    \draw[dashed] (82pt,8pt) -- (90pt,50pt);
    \draw[dashed] (100pt,8pt) -- (90pt,50pt);
    \draw[dashed] (136pt,8pt) -- (90pt,50pt);
    \draw[dashed] (118pt,8pt) -- (90pt,50pt);
    \filldraw[fill=black] (90pt,50pt) circle (1pt) node[above]{\scriptsize $Z_3Z_5Z_6Z_7Z_8$};
    \end{tikzpicture}
    }
    \qquad
    \subfloat[Sensitivity]{
    \begin{tikzpicture}[scale=0.7]
    \draw[->,>=stealth, thick] (0pt,-10pt) -- (0pt,80pt) node[above]{$E$};
    \filldraw[fill=black] (0pt,10pt) circle (1pt) node[left]{$-1$};
    \filldraw[fill=black] (0pt,50pt) circle (1pt) node[left]{$1$};
    \draw[thick] (20pt,10pt)-- (40pt,10pt);
    \draw[thick] (50pt,10pt)-- (70pt,10pt);
    \draw[thick] (80pt,10pt)-- (100pt,10pt);
    \draw[thick] (110pt,10pt)-- (130pt,10pt);
    \draw[thick] (140pt,10pt)-- (160pt,10pt);
    \draw[thick] (170pt,10pt)-- (190pt,10pt);
    \draw[thick] (20pt,50pt)-- (40pt,50pt);
    \draw[thick] (50pt,50pt)-- (70pt,50pt);
    \draw[thick] (80pt,50pt)-- (100pt,50pt);
    \draw[thick] (110pt,50pt)-- (130pt,50pt);
    \draw[thick] (140pt,50pt)-- (160pt,50pt);
    \draw[thick] (170pt,50pt)-- (190pt,50pt);
    \draw[->,>=stealth] (90pt,15pt) --++ (30pt,30pt);
    \draw[->,>=stealth] (90pt,15pt) --++ (0pt,30pt);
    \draw[->,>=stealth] (90pt,15pt) --++ (-30pt,30pt);
    \draw[->,>=stealth] (90pt,15pt) --++ (60pt,30pt);
    \end{tikzpicture}
    }
    \caption{(a)Degree corresponds to the range of the spin Hamiltonian; (b)Local sensitivity may be interpreted as the transition amplitude. }
    \label{fig:MBL}
\end{figure}
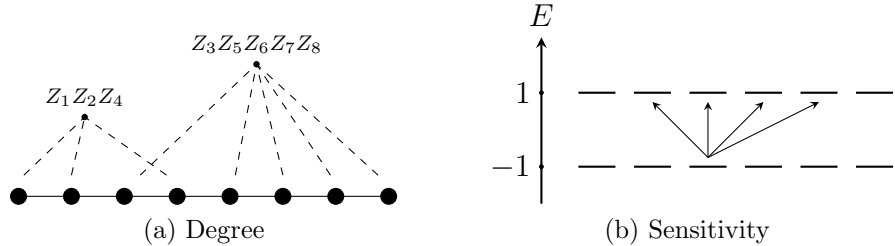

It is also interesting to make a physics analogy here. We can consider $f(s)$ as the ``energy" of the spin model. In other words, we could define a Hamiltonian 
\begin{equation}
    \hat{f}=\sum_{s\in \{0,1\}^n}f(s) |s\rangle \langle s|\,,
\end{equation}
which is diagonal in spin $Z$ basis. In general, we can expand $\hat{f}$ into monomials of $Z$:
\begin{equation}
    \hat{f}=a_0+\sum_j a_{1}^j Z_j+\sum_{j<k}a_{2}^{jk}Z_jZ_k+ \ldots \label{eq:expansion}
\end{equation}
as illustrated in Fig. \ref{fig:MBL}. The maximum number of $Z_j$ appearing in the expansion is named as the degree of $f(s)$. For example, the parity function $P(s)=\prod_{j=1}^n Z_j$ has degree $n$ 
 (the name degree here refers to the degree of a polynomial and should not be confused with the maximum degree of a subgraph). In physicists' langauge, the degree corresponds to the maximum size of operators in the Hamiltonian. 
Such Hamiltonians have been studied in  disordered spin systems and many-body localization\cite{nandkishore2015many}. The comparison is not very realistic since a Boolean function corresponds to a strange Hamiltonian with only two eigenvalues, and generically a lot of degenerate ground states. If we consider a perturbation such as $\lambda\sum_j X_j$ with small $\lambda$, the largest transition amplitude between the state with $f(s)=-1$ and those with $f(s)=1$ is determined by the sensitivity of $f$. In other words, sensitivity is related to the response of the spin model to perturbations. 

\begin{figure}[t]
    \centering
    \begin{tikzpicture}
    \filldraw[black,fill=gray, opacity=0.5] (0pt,70pt) rectangle ++(80pt,80pt);
    \filldraw[black, fill=white, opacity=1] (0pt,0pt) rectangle ++(80pt,60pt);
    \filldraw[black,fill=white, opacity=1] (100pt,90pt) rectangle ++(80pt,60pt);
    \filldraw[black, fill=gray, opacity=0.5] (100pt,0pt) rectangle ++(80pt,80pt);
    \node at (40pt,-20pt) {$P(s)=+1$};
    \node at (140pt,-20pt) {$P(s)=-1$};
    \node at (-40pt,30pt) {$f(s)=-1$};
    \node at (-40pt,120pt) {$f(s)=+1$};
    \node at (40pt,30pt){$\bar{H}^+$};
     \node at (140pt,40pt){${H}^-$};
      \node at (40pt,110pt){${H}^+$};
       \node at (140pt,120pt){$\bar{H}^-$};
       \filldraw [fill=black] (60pt,40pt) circle (2pt);
       \filldraw [fill=black] (60pt,80pt) circle (2pt);
        \filldraw [fill=black] (60pt,100pt) circle (2pt);
       \filldraw [fill=black] (115pt,70pt) circle (2pt);
       \filldraw [fill=black] (115pt,50pt) circle (2pt);
       \filldraw [fill=black] (115pt,100pt) circle (2pt);
       \filldraw [fill=black] (60pt,50pt) circle (2pt);
       \draw[thick] (60pt,50pt) -- (115pt,100pt);
        \draw[thick,blue, dashed] (60pt,40pt) -- (115pt,50pt);
        \draw[thick] (60pt,80pt) -- (115pt,70pt);
        \draw[thick,blue, dashed] (60pt,100pt) -- (115pt,100pt);
    \end{tikzpicture}
    \caption{Illustration of the relation between subgraph $H=H^+ \cup H^-$ (gray blocks) and function $f(s)$ (see text). This figure is a sketch of the hypercube $Q^n$, with the left two blocks representing even parity vertices, and the right two representing odd parity vertices. The links only connect points with opposite parity. 
    We let the blocks in the first row have $f(s)=1$ and the blocks in the second row have $f(s)=-1$. 
    Therefore according to the definition~\eqref{eqn: defn H}, the gray blocks belong to $H$ while the white blocks belong to the complement $\bar{H}$. The sensitivity of $f$ is related to the black (solid) links which are within $H$ or $\overline{H}$.
    } 
    
    \label{fig: H and f}
\end{figure}
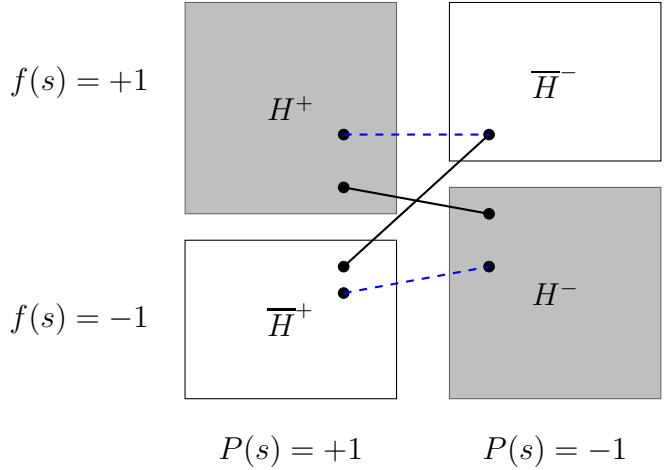

Without going further into the physical realization, we would like to see how sensitivity is related to the maximum degree of a certain subgraph. The relation is illustrated in Fig.~\ref{fig: H and f}. Since the hypercube $Q^n$ is bipartite,  i.e. it only contains edges between even parity and odd parity vertices (bit strings), for any Boolean function $f(s)$ one can define a partition of $Q^n$ into four parts, corresponding to $(f(s),P(s))=(\pm 1,\pm 1)$. Now we can recombine them and define
\begin{equation}
    H=\underbrace{\ke{s|f(s)=+1,P(s)=+1}}_{H^+}\cup\underbrace{\ke{s|f(s)=-1,P(s)=-1}}_{H^-}\,.
\label{eqn: defn H}
\end{equation}

In Fig.~\ref{fig: H and f}, $H$ is the union of two gray blocks. In general a vertex $x$ in $H^+$ (parallel discussion applies to other blocks as well) is linked to $n$ vertices in $\bar{H}^-\cup H^-$, the local sensitivity $s(f,x)$ is the number of links from $x$ to the vertices in $H^-$ only, since they have the opposite $f$ value. These links are exactly the internal edges of $H$ that connect to $x$, the number of which is the degree of $x$ in $H$. The sensitivity is a maximum of the local sensitivity over all vertices, both in $H$ and $\bar{H}$ and corresponds to the maximum degree of $H$ or $\overline{H}$, whichever is larger:
\begin{equation}
    s(f)={\rm max}\left( \Delta(H),\Delta(\bar{H}) \right) \,.
\end{equation}
Compared with the discussion in section~\ref{sec: tight} one can see that the Boolean function $h(s)$ defined in Eq.~\eqref{eqn: g and h} that has been used in the construction of $H$ by Chung, Furedi, Graham and Seymour, has sensitivty $\sqrt{n}$.

If the size of the subgraph $|H|\neq 2^{n-1}$, one of $H$ and $\overline{H}$ will have a size of more than half of $Q^n$. Applying Huang's theorem, we will obtain the bound $s(f)\geq \sqrt{n}$. Apparently, this cannot be true for all Boolean functions, since we can easily construct Boolean functions with sensitivity as low as $0$ (e.g. constant function $f(s)=1$) or $1$ (e.g. $f(s)=Z_1$). That must mean that for such functions, the construction here gives an $H$ of size exactly $2^{n-1}$. 
Indeed, the size of $H$ can be determined in a formula similar to the discussion in section~\ref{subsec: CFGS example}. Define
\begin{equation}
    g(s)=f(s)P(s)\,,
\end{equation}
in the same way as Eq.~(\ref{eqn: g and h}), we see that $H$ is defined by $g(s)=1$. Thus, 
\begin{equation}
    |H|=\frac{1}{2} \Tr\left( 1+g(s)\right)=2^{n-1}+ \frac{1}{2} \Tr \left( f(s)P(s) \right) \,.
    \label{eq:size of H}
\end{equation}
Therefore if $f(s)$ does not contain the parity $P(s)$,  i.e. if its degree is less than $n$, $|H|=2^{n-1}$ and we could not obtain a bound on sensitivity of $f$ in the construction above.

This problem can be resolved if we simply replace the parity function $P(s)$ in the discussion above by the highest degree term in the expansion of $f(s)$ in Eq.~(\ref{eq:expansion}). When $f(s)$ has degree $m$, without loss of generality, we can assume one of the highest order term in the expansion of $f(s)$ is 
\begin{equation}
    P_m\equiv Z_1Z_2...Z_m \,.
\end{equation}
$P_m(s)$ is the parity function on the lower dimensional hypercube, an $m$-dimensional ``surface" of $Q^n$. The position of the ``surface" is determined by the remaining $n-m$ coordinates. For example, we can consider a sub-hypercube $Q_m$ consisting of strings of the form $s=\ke{s_1s_2...s_m,00...0}$ with arbitrary $s_1,s_2,...,s_m$. For any such choice, function $f(s)$ can be restricted to $Q_m$ to define a function on the $m$-dimensional hypercube. Obviously, the restriction does not increase the senstivity of the function. By construction, this induced function has maximum degree $m$. Applying Eq.~(\ref{eq:size of H}) to the subgraph $Q_m$, we find a subgraph $H_m\subseteq Q_m$ with size different from $2^{m-1}$. Therefore $s(f)\geq s_m(f)\geq\sqrt{m}$. Here we denote $s_m(f)$ as the sensitivity of $f$ when restricted to the subgraph. 
 In summary, we see that combining Huang's theorem and the result of \cite{gotsman1992equivalence} leads to
\begin{equation}
    s(f)\geq\sqrt{\deg(f)} \,.
\end{equation}

Finally, to explain the sensitivity conjecture one needs to introduce the concept of block sensitivity $bs(f)$, which may be thought as a generalization of the sensitivity by allowing multiple flips at once. In the spin language, this corresponds to considering a perturbation of multiple spin flips, of the form $X_1X_2..X_m$. The definition of block sensitivity is given as follows:

\begin{defn}[Block sensitivity.] 
For a Boolean function $f$: $\{0,1\}^n \rightarrow \{ -1,1\}$, the local block sensitivity $bs(f,x)$ on the input $x$ is defined as the maximum number of disjoint blocks $b_1,b_2,\ldots,b_k$ of $\{1,2,\ldots,n\}$ such that for each $b_j$, $f(x)=-f(x')$ where $x'$ and $x$ differ in exactly all the coordinates $x_{i}$ with $i \in b_j$. The global block sensitivity $bs(f)=\max_x bs(f,x)$ is defined by maximizing over all inputs $x$. Note that $bs(f)\geq s(f)$ following the definition. 
\end{defn}
Nisan and Szegedy\cite{nisan1994degree} showed that the block sensitivity is upper bounded by polynomial of $\deg(f)$, more exactly $bs(f)\leq 2\deg(f)^2$. Combining with the inequality $s(f)\geq \sqrt{\deg(f)}$, Huang achieved the proof of sensitivity conjecture in the form $
bs(f)\leq 2s(f)^4$. We won't make further discussion on block sensitivity and its relation to degree in this note.

\section{Further discussion}

It is interesting to see if the connection of Huang's proof with physics of spin chains and fermions helps for the understanding of any related math problem, such as the generalization from functions on $Q^n$ to other graphs. It is also interesting to notice the relation between degree of Boolean functions and the concept of operator size, which is a new criteria proposed for many-body quantum chaos\cite{roberts2015localized,hosur2016characterizing}. It is natural to ask whether the relation between degree and sensitivity can be applied to define new measure of chaos and complexity in many-body systems.

\section*{Acknowledgement}

We would like to dedicate this note to our late colleague and mentor Shou-Cheng Zhang, who had always encouraged us to explore and appreciate the simplicity and universality of all human knowledge. Y.~G. is supported by the Gordon and Betty Moore Foundation EPiQS Initiative through Grant (GBMF-4306) and DOE grant, DE-SC0019030. X.-L.~Q. is supported by the National Science Foundation under grant No.~1720504, and the Simons Foundation.

\bibliography{refs}

\end{document}